\documentclass[aps,prb,twocolumn]{revtex4-1}

\usepackage{amsfonts} 
\usepackage{amsmath,amssymb,graphicx}
\usepackage{color,times}
\usepackage{siunitx}
\usepackage{xcolor}
\definecolor{darkblue}{rgb}{0, 0, 0.8}
\definecolor{ForestGreen}{RGB}{34,139,34}
\definecolor{Reveiwer1edits}{RGB}{255,30,30}
\definecolor{Reveiwer2edits}{RGB}{10,255,10}

\usepackage[colorlinks=true, breaklinks=true, linkcolor=darkblue, citecolor=darkblue, urlcolor=darkblue]{hyperref}
\newcommand{\doilink}[2]{\href{http://dx.doi.org/#1}{#2}}

\begin{document}
\title{A basic introduction to ultrastable optical cavities for laser stabilization}

\author{Jamie A. Boyd} 
\author{Thierry Lahaye} 
\affiliation{Universit\'e Paris-Saclay, Institut d'Optique Graduate School,\\
CNRS, Laboratoire Charles Fabry, 91127 Palaiseau Cedex, France}

\date{\today}

\begin{abstract}
We give a simple introduction to the properties and use of ultrastable optical cavities, which are increasingly common in atomic and molecular physics laboratories for stabilizing the frequency of lasers to linewidths at the kHz level or below. Although the physics of Fabry-Perot interferometers is part of standard optics curricula, the specificities of ultrastable optical cavities, such as their high finesse, fixed length, and the need to operate under vacuum, can make their use appear relatively challenging to newcomers.  Our aim in this work is to bridge the gap between generic knowledge about Fabry-Perot resonators and the specialized literature about ultrastable cavities. The intended audience includes students setting up an ultrastable cavity in a research laboratory for the first time and instructors designing advanced laboratory courses on optics and laser stabilization techniques. 
\end{abstract}

\maketitle

\section{Introduction}
Over the past two decades, optical clocks have revolutionized frequency standards, now achieving fractional accuracies in the $10^{-18}$ range.\cite{bothwell2019,brewer2019,zheng2022,khabarova2022} Crucial to reaching such performance are interrogation lasers whose frequencies are stabilized through the use of high-finesse, ultrastable cavities. Similar cavities are also used for gravitational wave interferometers.\cite{aasi2015} More recently, beyond the field of metrology, these ultrastable cavities have become increasingly widespread in atomic and molecular physics laboratories for addressing narrow transitions. Cavities based on ultralow expansion (ULE) glass spacers, which once required complex custom developments, are now commercially available, making it relatively easy to reach linewidths of a few kilohertz (corresponding to a quality factor $Q\sim 10^{11}$). These cavities facilitate research in a wide range of areas, including quantum gases, molecular physics, and R
ydberg atoms.

Although the physics of Fabry-Perot interferometers is commonly taught in undergraduate optics curricula, the distinct characteristics of such ultrastable cavities (high finesse in the tens of thousands, operation under vacuum, choice of spacer material, temperature stability, importance of realizing a proper mode-matching) are rarely covered, leading to a knowledge gap for students and researchers when it comes to setting one up for the first time. In this article, we aim to provide a concise introduction to the physics of ultrastable, high-finesse cavities, with the goal of bridging the gap between general knowledge of Fabry-Perot resonators and the practical use of an ultrastable cavity in a laboratory setting. For more detailed introductions along similar lines, we refer readers to Refs.~\onlinecite{martin2011} and \onlinecite{fox2002}. 

The article is structured as follows. In the first part, we provide background on the practical requirements of ultrastable cavities to achieve ultralow fluctuations of the resonance frequencies. In the second part, we explain how to set up an ultrastable cavity in practice, including mode-matching, implementing a Pound-Drever-Hall lock, and measuring the finesse of the cavity. We conclude with a discussion on why we believe that including such a setup in an advanced instructional laboratory course can lead to a variety of interesting experiments focusing on various aspects of optics for undergraduate students.

\section{Conceptual Background}
The ultimate goal of locking a laser to an optical cavity is to minimize its frequency fluctuations. Laser frequency fluctuations are primarily due to technical noise in the laser, such as temperature fluctuations, mechanical vibrations, or fluctuations in the current in the case of a semiconductor laser. Such noise, in combination with the fundamental quantum noise, contributes to increase the laser linewidth. Depending on the source of noise, the fluctuations take place on different time scales: ``fast'' fluctuations, which occur on the order of tens of microseconds; low-frequency jitter, on timescales of milliseconds to seconds; and slow drifts of the central laser frequency over the course of hours, days, and weeks, due primarily to thermal effects.\cite{fox2002, riehle2003}  

A cavity helps minimize laser frequency fluctuations primarily on the first two time scales by serving as a precise, stable reference. The precision of the cavity is determined by its resonance linewidth; narrow cavity linewidths help reduce the fast fluctuations and jitter in the laser frequency. The stability of the cavity determines the changes in the center frequency of the resonances over time; high stability allows one to reduce slow drifts of the laser's central frequency. Ultimately, slow drifts are accounted for by referencing the stabilized laser to an atomic transition. 

In this section, we start by a review of the basic physics of optical cavities, including the relationship between linewidth and finesse, and then turn our attention to the specific requirements for designing an ultrastable cavity. 

\subsection{Brief review of optical cavities}
\subsubsection{Linewidth and finesse}
A linear optical cavity consists of two mirrors whose reflective surfaces face each other (see Fig. \ref{figure1}a). Monochromatic light entering the cavity will be resonant when the round-trip distance $2L$ is an integer multiple $p$ of the light's wavelength $\lambda$. For a lossless cavity on resonance, the incoming light is fully transmitted through the cavity, with none reflected (intensity reflection coefficient $R=0$), due to constructive interference of the field circulating within the cavity and destructive interference between the reflected field and the field leaking out the first mirror. Off resonance, almost all light is reflected and none transmitted, due to destructive interference in the cavity.

The frequency spacing between two successive resonances is called the free spectral range (FSR) and is determined by the length of the cavity: 
\begin{equation} \label{eq:FSR}
    \nu_{\mathrm{FSR}} = \frac{c/n}{2L},
\end{equation}
where $c$ is the speed of light in a vacuum, $n$ the refractive index of the medium between the two mirrors, and $L$ is the length of the cavity. The finesse of the cavity is defined as the ratio of the free spectral range to the full-width half-maximum linewidth of the resonance peaks, denoted $\delta \nu$, and is determined by the reflectivity of the mirrors:\cite{riehle2003}
\begin{equation} \label{eq:finesse_def}
    \mathcal{F} = \frac{\nu_{\mathrm{FSR}}}{\delta \nu} = \frac{\pi \sqrt{r_1 r_2}}{1-r_1 r_2}.
\end{equation}
Here, $r_1$ and $r_2$ are the electric field reflection coefficients of the two mirrors. From Eq. (\ref{eq:finesse_def}), one can see that the higher the reflectivity of the mirrors, the higher the finesse. For example, $|r_1|=|r_2|=98.4\%$ results in a finesse of 100, while $|r_1|=|r_2|=99.992\%$ yields a finesse of 20,000. 

Once the finesse of the cavity is fixed, the choice of cavity length must take into account several factors: the desired free spectral range, the resulting linewidth of the resonances, and the practical space considerations. Commonly, ultrastable cavities used in atomic and molecular physics laboratories are on the order of 10 cm long, giving $\nu_{\mathrm{FSR}}\simeq 1.5$~GHz. For this free spectral range, a finesse of 100 gives $\delta \nu \simeq 15$ MHz, whereas a finesse of 20,000 gives $\delta \nu \simeq 75$ kHz. Since the achievable narrowing of a laser's linewidth is a fraction of the linewidth of the cavity's resonance, high finesse is an important requirement for designing cavities that can address narrow atomic and molecular transitions.

In the time domain, the ideal linewidth of the resonances is directly related to the $1/e$ decay time $\tau$ of the intensity in the cavity: $\delta \nu = 1/(2 \pi \tau)$, giving
\begin{equation} \label{eq:finesse_in_terms_of_tau}
    \mathcal{F} = 2 \pi \tau \nu_{\rm FSR}.
\end{equation}
A long cavity decay time corresponds to a high finesse and a narrow linewidth on resonance. Measuring $\tau$ provides an accurate way to measure the finesse of the cavity of a known length $L$ (see Sec. \ref{sec: ringdown finesse measurement}).

\subsubsection{Higher-order modes} \label{subsec:Higher-order modes}

\begin{figure}
    \centering
    \includegraphics[width=0.48\textwidth]{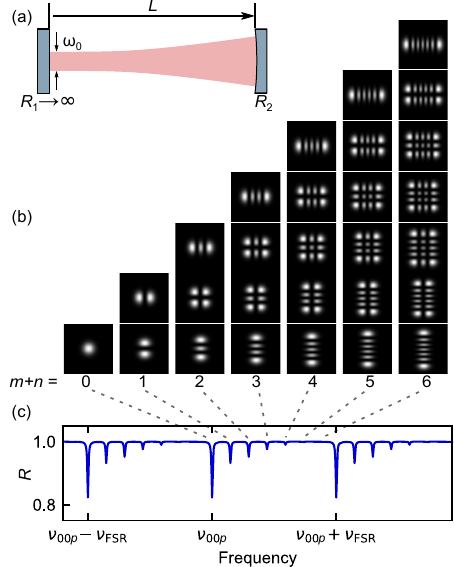}
    \caption{(a)~Diagram of a planoconvex optical cavity, where $R_1$ and $R_2$ denote the radii of curvature of the two mirrors. (b)~Calculated Gauss-Hermite intensity profiles for each transverse mode. For a perfectly cylindrical cavity, modes with the same value of $m+n$ are degenerate in frequency, and the transverse intensity profile in transmission can be a superposition of all transverse modes with the same value of $m+n$. (c) Calculated signal in reflection while scanning the laser frequency slowly over three free spectral ranges, where  $R$ represents the intensity reflection coefficient and $\nu_{00p}$ is the frequency of a fundamental transverse mode resonance. Each dip corresponds to a cavity resonance, with the deepest dip corresponding to the fundamental mode. The reflection signal was calculated for a cavity with a finesse of 100 to portray the non-zero linewidth of the resonances and a ratio $L/R_{2}=1/5$, and the relative amplitudes were set to match the experimental results in Fig. \ref{figure5}(b).}
    \label{figure1}
\end{figure}

Thus far, we have considered the resonances of an optical cavity as depending only on the longitudinal mode of the incoming light. This simplification is valid when considering only the fundamental transverse mode: a pure Gaussian beam. In reality, however, imperfect coupling of a laser into a cavity leads to the excitation of higher-order transverse modes, whose resonance frequencies differ from that of the fundamental mode. These higher-order transverse modes are typically of the Gauss-Hermite family and are denoted by two integers, $m$ and $n$, for the two transverse directions [Figs.~\ref{figure1}(b) and \ref{figure1}(c)]. Their wave amplitudes are given by the product of a Gaussian beam expression, two Hermite polynomials corresponding to the two transverse directions, and a phase term, referred to as the Gouy phase:
\begin{equation} \label{eq:phase}
\text{exp}\left[-i(m+n+1)\text{arctan}(z/z_R)\right],    
\end{equation}
where $z_R$ is the Rayleigh range~\cite{brooker2002} (see Appendix \ref{higher order Gauss-Hermite} for the full expression of the wave amplitude). 

For a mode to be resonant in the cavity, the phase accumulated during one round trip must be a multiple of $2\pi$. Equation (\ref{eq:phase}) therefore determines the resonance condition, from which one can derive the resonance frequencies to be:
\begin{equation}\label{eq:nu_mnp}
    \frac{\nu_{mnp}}{\nu_{\mathrm{FSR}} } = p + \frac{(m+n+1)}{\pi} \cos^{-1}\sqrt{\left(1-\frac{L}{R_1}\right)\left(1-\frac{L}{R_2}\right)},
\end{equation} 
where $p$ is an integer labeling the longitudinal mode, $L$ is the length of the cavity, and $R_1$ and $R_2$ are the radii of curvature of the two mirrors.\cite{kogelnik1966} The phase factor given by Eq.~(\ref{eq:phase}) lifts the frequency degeneracy of transverse modes with different values of $m+n$. 

From Eq. (\ref{eq:nu_mnp}), one can see that for a given transverse mode with fixed values for $n$ and $m$, the resonance frequencies of the corresponding longitudinal modes are spaced by the free spectral range of the cavity. For a given longitudinal mode $p$, the spacing depends on the ratio between the length of the cavity and the radius of curvature of the mirrors. With all else held constant, increasing the ratio $L/R_{1,2}$ increases the frequency spacing between corresponding transverse modes. 

The need for narrow linewidth resonances therefore relates to the choice of mirror geometry for the cavity and explains why, despite its frequent use in scanning Fabry-Perot cavities, a confocal geometry ($L=R_1=R_2$) is not optimal for applications where high-finesse is crucial. In a confocal geometry, the spacing between two consecutive resonances is given by $\nu_{\mathrm{FSR}}/2$, so all even higher-order modes are degenerate with the fundamental mode. However, this degeneracy is never perfect; higher-order modes will appear as many small peaks close to the fundamental resonance. These peaks increase the linewidth of the resonance and reduce the effective finesse of the cavity. Therefore, when designing high-finesse cavities for locking lasers, one aims to space the resonances of the higher-order modes sufficiently far away from the fundamental mode. In addition, one wants to choose a spacing such that the higher-order transverse modes do not overlap with any fundamental 
transverse modes (of higher longitudinal modes). It is common to choose a planoconvex cavity with a ratio $L/R_2$ that gives a frequency spacing between consecutive higher-order transverse modes on the order of 100 MHz and a high least common multiple with the free spectral range. 

Figure \ref{figure1}(c) shows an example of the calculated intensity reflected from a cavity with a ratio $L/R_2 = 1/5$ when scanning the laser over three free spectral ranges \cite{footnote0}. The higher-order modes are equally spaced with decreasing amplitudes away from the fundamental mode. Calculated images of the intensity patterns of the transverse modes are pictured in Fig.~\ref{figure1}(b). Assuming the cavity has perfect cylindrical symmetry, transverse modes with the same value of $m+n$ are degenerate in frequency, and the intensity pattern viewed in transmission through the cavity is a superposition of all modes with the same value $m+n$. In practice, however, no cavity has perfect cylindrical symmetry due to minute imperfections of the mirrors. These imperfections lift the degeneracy of the transverse modes, and the corresponding frequency splitting can be measured for cavities with sufficiently high finesse.\cite{fabre1986}

\subsection{Design requirements for ultrastable cavities}
In addition to having a high finesse and optimally spaced higher-order modes, a cavity for locking lasers should be ultrastable to avoid slow drifts of the laser frequency. Ideally, the resonance frequency of the cavity should drift as little as possible. Practically, it is relatively easy to achieve variations of less than several hundred kHz over the course of one day using a ULE spacer. As shown in Eqs. (\ref{eq:FSR}) and (\ref{eq:nu_mnp}), the cavity's resonance frequencies depend primarily on its length and the index of refraction of the medium between the mirrors.\cite{footnote1} The relative variation in resonance frequency then fulfills

\begin{equation}
    \frac{\delta \nu}{\nu} \leq \frac{\delta n}{n} + \frac{\delta L}{L}.
\end{equation}
To achieve $\delta \nu < 1$ MHz at optical frequencies, the relative fluctuations $\delta \nu/\nu$ should be less than $10^{-9}$. 

In the following sections, we explain how cavities are designed to reach such stunning stabilities. 

\subsubsection{Operating under vacuum}

The index of refraction $n$ of air at room temperature and pressure is on the order of $n-1\simeq3 \times 10^{-4}$. However, atmospheric pressure depends on many factors and can vary by up to 10\% over the course of several days. Thus, with ambient air between a cavity's mirrors, the resulting relative fluctuations $\delta n/n$ would be on the order of $10^{-5}$, yielding an unacceptably large value for $\delta \nu/\nu$. 
To eliminate these fluctuations, the space in between the two cavity mirrors needs to be evacuated. A moderate vacuum, with a residual pressure below $10^{-3}$ mbar, reduce the relative fluctuations of the index of refraction sufficiently.

In practice, the choice of the vacuum pump is determined by the requirement that it introduce no vibrations to the system. Pumps that operate using spinning parts are therefore not good options. Instead, ion pumps are typically used as they have no moving parts. Ion pumps easily maintain pressures around $10^{-7}$~mbar without bakeout. In these conditions, the relative fluctuations of the refractive index are on the order of $10^{-14}$ and are thus entirely negligible.

\subsubsection{Minimizing length fluctuations}

\begin{figure}[t]
    \centering
    \includegraphics[width=85mm]{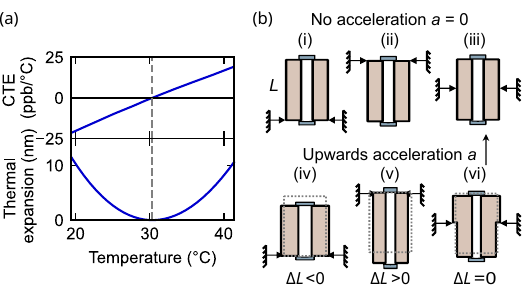}
    \caption{(a) Coefficient of thermal expansion (CTE) and thermal expansion vs. temperature for a batch of ultralow expansion (ULE) glass with a zero-crossing temperature $T_{\rm ZC}=30.4$ °C. At $T_{\rm ZC}$, the relative length fluctuations for a spacer of length 10 cm are on the order of $10^{-14}$ for temperature fluctuations of $\pm 10$ mK. (b) A simple setting illustrating the influence of vibrations on the length of a cavity, and how this can  be mitigated by a proper choice of the suspension. A vertical cavity, of length $L$ in the absence of acceleration, is either supported from the bottom (i), suspended from the top (ii), or suspended mid-point (iii). When the cavity undergoes an upwards acceleration $a$, its length decreases in the first case (iv) and increases in the second case (v). For the mid-point suspension, the upper part of the cavity gets compressed, while the lower part gets elongated, resulting in an overall unchanged cavity length. }
    \label{figure2}
\end{figure}

\begin{figure*}
    \centering
    \includegraphics[width=\textwidth]{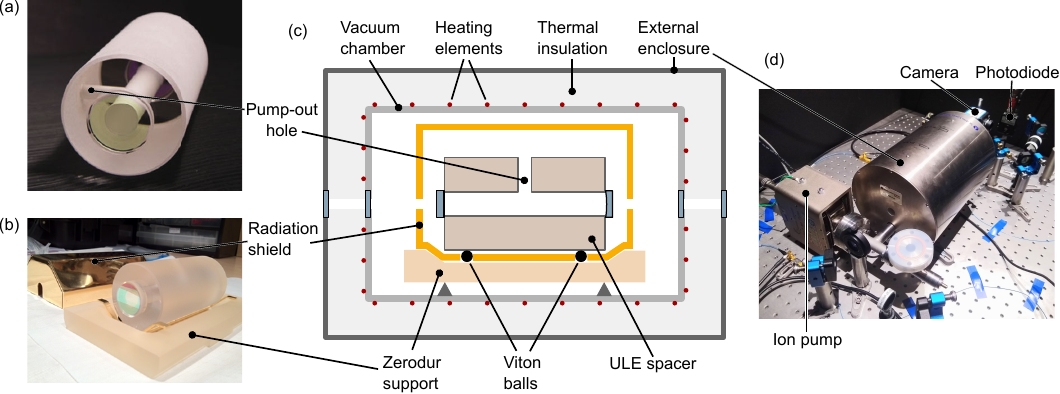}
    \caption{(a) Photo showing the ULE glass spacer and mirrors of an ultrastable cavity (image courtesy of Stable Laser Systems). The pump-out hole for evacuating the cavity is visible on the left. (b) Photo of our SLS cavity showing both the Zerodur support and the radiation shield. (c) Diagram of the cross-section of an SLS ultrastable cavity. The ULE glass spacer is supported by fluororubber balls for thermal and vibrational insulation and surrounded by a radiation shield. The radiation shield rests on a support in Zerodur, which is supported on the walls of the vacuum chamber. Around the vacuum chamber, heating elements maintain the inside at a constant temperature. The external enclosure surrounds thermal insulation around the vacuum chamber. (d) Photo of our setup showing the external enclosure of the cavity connected to the ion pump as well as the camera and photodiode for measuring the signal in transmission.}
    \label{figure3}
\end{figure*}

The remaining frequency fluctuations arise primarily from variations in the distance between the two mirrors. The mirrors are fixed to the ends of a spacer of solid material (see Fig.~\ref{figure3}) so that the distance between them depends on variations in the length of the spacer. Such length variations are due primarily to changes in temperature. To first order, the relative change in length of a material due to changes in temperature is given by
\begin{equation}
    \frac{\delta L}{L} = \alpha(T) \, \delta T,
\end{equation}
where $\alpha(T)$ is the coefficient of thermal expansion (CTE) of the material. To minimize length fluctuations, spacers are made of a material with the lowest possible CTE near room temperature. 
The most common choice of material is ultralow expansion (ULE) glass, which is a compound of silica and titanium dioxide, as its CTE is very low near room temperature and even vanishes at a so-called \emph{zero-crossing temperature} $T_{\rm ZC}$. The CTE and exact value of $T_{\rm ZC}$ is empirically determined for each ULE spacer, as it depends not only on the composition of the material but also on its fabrication conditions, which can vary between each production cycle. 

Figure \ref{figure2}(a) shows the CTE of a particular batch of ULE with a zero-crossing temperature at $30.4\,^\circ$C. Usually, the zero-crossing temperature is chosen to be slightly above room temperature (typically around $30-35 \,^\circ$C) as it is easier to maintain a constant temperature by heating than by cooling. Resistive heaters encircle the vacuum chamber containing the cavity, as shown in Fig.~\ref{figure3}(c), maintaining temperature stability better than $\pm 10$ mK. As can be inferred from Fig.~\ref{figure2}(a), such temperature fluctuations around the zero-crossing temperature induce relative length fluctuations for a spacer of 10 cm on the order of $10^{-14}$. Even at a deviation of $\pm 0.5$ $^{\circ}$C from the zero-crossing temperature, the relative length fluctuations are on the order of $10^{-11}$ and are thus negligible.

\subsubsection{Minimizing the effects of vibrations}

Mechanical vibrations due to acoustic or seismic noise can be another significant source of fluctuations of the cavity length. For applications where one does not need to address extremely narrow transitions (below the kilohertz range), the effects of such vibrations can usually be neglected. However, it is crucial to minimize their effect to reach state-of-the-art performance for gravitational wave detectors~\cite{aasi2015} or for locking ultrastable lasers used in the operation of optical clocks.\cite{martin2011} This is achieved by (i) using vibration-isolation platforms on which the reference cavity is mounted, and (ii) using a clever shaping and mechanical suspension of the cavity that minimizes the influence of a given level of vibrations on the cavity length. Several approaches exist, ultimately relying on finite-element modeling of the cavity response to accelerations.\cite{millo2009} In this section, we illustrate this type of vibration mitigation on a conceptually s
imple case, namely that of a midpoint vertically suspended cavity, a configuration often used in actual setups to minimize sensitivity to vertical vibrations. 

Figure~\ref{figure2}(b) shows three possible ways to hold a vertical cavity spacer of length $L$: it can (i) rest on the bottom, (ii) be hanging from the top, or (iii) be suspended midpoint. Now suppose the cavity is subjected to vertical vibrations. For frequencies much lower than the mechanical resonance frequency of the  spacer, the change $\Delta L$ of the cavity length due to an upwards acceleration $a$ can be calculated in the quasistatic case. In case (i), the cavity length \emph{decreases} by a quantity which can be shown to be $\Delta L = a\rho L^2/(2E) $, where $\rho$ is the density of the spacer material and $E$ its Young's modulus [Fig.~\ref{figure2}(b), panel (iv)].  For the case of a spacer hanging from the top, the same acceleration would produce an \emph{increase} in length [Fig.~\ref{figure2}(b), panel (v)], by the same quantity. Thus, if the cavity is suspended in the middle, its top part contracts while its lower part extends by the same length, resulting i
n a vanishing net change in the spacer length [Fig.~\ref{figure2}(b), panel (vi)].  

\subsubsection{Recent developments in ultrastable cavity design}
Another important effect that impacts the stability of a cavity is thermal noise. Even for a perfectly stable temperature $T$, the spacer, mirror substrates, and mirror coatings undergo Brownian fluctuations that lead to noise in the effective cavity length scaling as $\sqrt{T}$. For state-of-the-art optical clocks, this noise is a major limitation. The current strategy to circumvent this problem is to use a cryogenic cavity with a spacer made of a single crystal silicon held at its zero-crossing temperature of 16~K.\cite{wiens2014, kedar2023} This approach, which is obviously quite involved technically, has two main advantages: (i) at cryogenic temperatures, thermal fluctuations are drastically reduced relative to those at room temperature; and (ii) using a single crystal nearly eliminates longterm length drift (``aging") of the spacer, due to the crystal's almost-perfect lattice structure.

\section{Ultrastable cavities in practice}

We now turn to the practical implementation of a room-temperature ultrastable ULE cavity for locking a 780~nm external cavity diode laser (ECDL). In this work, we use an ultrastable cavity from Stable Laser Systems (SLS) that we previously used for two-photon Rydberg excitation of single Rb atoms.\cite{beguin2013} Since this application required laser linewidths only on the order of tens of kilohertz, our cavity is horizontally rather than vertically suspended and kept near room temperature. It has a nominal finesse of 25,200 at 780 nm, and a free spectral range of 1.5 GHz.

\subsection{Experimental setup} \label{sec: experimental setup}

Figure~\ref{figure3} shows a cross-sectional diagram and photos of our cavity. A cylindrical  spacer made of ULE glass separates the two cavity mirrors fixed on either end. A long, cylindrical hole is bored through the center of the spacer and maintained under vacuum via a pump-out hole. The spacer rests on four Viton balls supported by a V-shaped Zerodur support. The viton balls minimize both vibrational and thermal couplings to the external environment. A radiation shield surrounds the spacer, resting on the Zerodur support, to minimize radiative heat transfer. This entire ensemble is enclosed in a vacuum chamber pumped typically to  $10^{-7}$~mbar using a $10$~L$/$s ion pump.  The Zerodur support rests on the inner walls of the vacuum chamber. The whole chamber is maintained at the zero-crossing temperature of our spacer (34.49 $^{\circ}$C) using resistive heating elements and a PID controller (Wavelength Electronics, LFI-3751). Outside the vacuum chamber is further therma
l insulation, surrounded by a final external enclosure.

\subsection{Mode matching} \label{sec: mode matching}
Our cavity is planoconvex, as depicted in Fig.~\ref{figure1}(a), with a radius of curvature $R_2=500$~mm and length $L=100$~mm. To mode-match the incoming beam with the fundamental Gaussian mode of the cavity, we must match the waist size and position of the incoming laser beam to those of the cavity's fundamental mode. Knowing $R_2$ and $L$, we can calculate the Rayleigh range $z_R$ of the fundamental Gaussian mode of the cavity. The radius of curvature of the Gaussian wavefront, given by 
\begin{equation}
    R(z) = z + \frac{z_R^2}{z},
\end{equation}
where $z$ denotes the distance from the waist, must match the mirrors' radii of curvature. Thus, the waist position $z=0$ coincides with that of the flat mirror, and setting $R(L)=R_2$ determines the Rayleigh range:
\begin{equation} \label{rayleigh range}
    z_R = \sqrt{L(R_2-L)}.
\end{equation}
The $1/e^2$ radius is then given by 
\begin{equation} \label{omega_0}
    w_0 = \sqrt{\frac{z_R \lambda}{\pi}}, 
\end{equation}
where $\lambda$ is the laser wavelength. Combining Eqs. (\ref{rayleigh range}) and (\ref{omega_0}) for 780 nm gives the $1/e^2$ radius of our cavity's fundamental mode as 223~µm. 

\begin{figure}[t]
    \centering
    \includegraphics[width=0.42\textwidth]{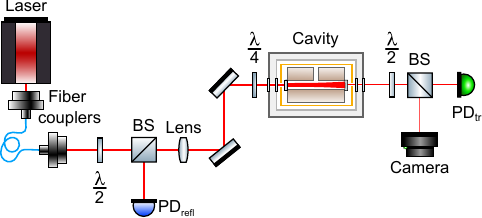}
    \caption{Simple optical setup for coupling light into a cavity with photodiodes in reflection and transmission. In our setup, we use a lens with $f=150$ mm for mode matching.}
    \label{figure4}
\end{figure}

The almost-perfectly Gaussian beam coming out of the fiber collimator (Thorlabs, CFC-2-B), with its own waist size and position,  needs to be matched to the cavity mode using an appropriately placed lens (or combination of lenses). Using Gaussian optics with ABCD matrices\cite{yariv1989, brooker2002} or an online Gaussian beam calculator,\cite{footnote2} one can find the optimal combination of fiber collimator position, lens position, and lens focal length. For our setup, we measure that the waist almost coincides with the output of the fiber collimator, with a $1/e^2$ radius of 100~$\mu$m. We then calculate that using a single lens of $f=150$~mm positioned at 204~mm from the collimator and 421~mm from the plane cavity mirror is a good solution, as it gives a reasonably compact setup. 

As shown in Fig. \ref{figure4}, we also use a $\lambda/2$ waveplate and beamsplitter before the cavity to reduce the power of the incident laser beam. It is important to have low power at the input of the cavity since, on resonance, the field circulating in the cavity has an intensity orders of magnitude greater than the incident field, given by $I_{\rm cav}/I_{\rm inc} = \mathcal{F}/\pi$. For our cavity, the intracavity field is approximately $8,000$ times the incident power, so we use an incident beam of $\sim 100$ µW to give $<1$ W within the cavity. Too much intracavity power can temporarily deform, or even permanently damage, the cavity mirrors. 

\begin{figure}
    \centering
    \includegraphics[width=0.53\textwidth]{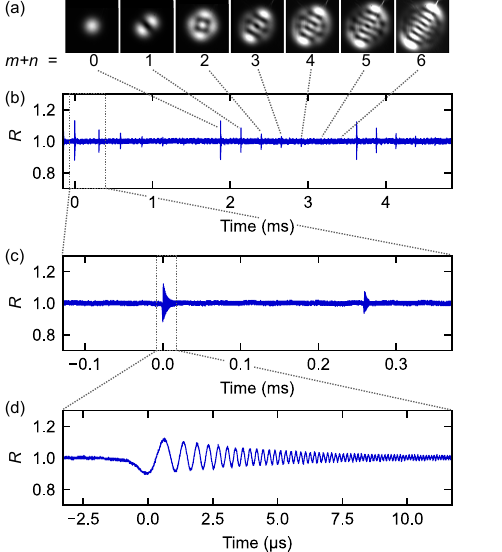}
    \caption{Measured reflection coefficient $R$ from cavity while scanning the laser frequency quickly over approximately three free spectral ranges. (a) Cavity modes imaged in transmission through the cavity, where $m$ and $n$ denote the integers labelling the transverse modes TEM$_{m,n}$. The gain and exposure time of the camera were increased to image each successive mode, so the relative intensity is not comparable between images. (b) Full scan over three FSR. Here, the coupling into the cavity was purposely not optimized into the fundamental mode to enhance the visibility of the higher-order modes.  In (c), we decrease the time scale by a factor of approximately 10 and center on the first two modes depicted in (b). Here, it starts to become noticeable that the resonance dips and peaks have non-zero linewidths. From (c) to (d), we decrease the time scale by a factor of approximately 30 and center on the fundamental mode. The reflection signal decreases on resonance, here to approximately 85\% of complete reflection. It then increases past 115\% reflection and oscillates at an increasing frequency with an amplitude that decays exponentially. Here, the cavity resonance FWHM is $\delta \nu \simeq 60$ kHz, corresponding to a $1/e$ decay time of $\tau = 2.7$ µs, and we scan the laser frequency at a rate of approximately 900 MHz$/$ms.}
    \label{figure5}
\end{figure}

Aligning the laser into the cavity from scratch is challenging, so it is helpful to use multiple diagnostic tools. We use a combination of a camera in transmission and photodiode in reflection, as shown in Fig. \ref{figure4}, to find and optimize the signal. While scanning the laser frequency over more than one FSR, we adjust the alignment of the incoming beam into the cavity using the mirrors until we see a transmission signal on the camera, as in Fig. \ref{figure5}(a). Scanning the laser frequency relatively slowly ($1-5$ Hz) helps us to distinguish a genuine signal from background reflections, as a signal flashes on the camera at a rate proportional to the scan frequency. After finding a signal, we align the photodiode in reflection. While scanning the laser frequency, the signal on the reflection photodiode should resemble that of Fig. \ref{figure5}(b). The goal is to maximize the coupling into the cavity for the fundamental Gaussian mode. To do so, we use the camera and 
reduce the scan amplitude until the fundamental mode is the only one flashing on the camera. Then, we maximize the depth of the corresponding dip on the reflection photodiode relative to the fully reflected signal, by adjusting the coupling mirrors.

The process of coupling is iterative: if the signal after aligning the mirrors is still sub-optimal, we slightly adjust the position of the lens and restart the process. We aim for the reflection dip to be $\gtrsim80$\% of the fully reflected signal while scanning the laser and for the higher-order modes to be nearly nonexistent. Decreasing the scan speed helps one achieve deeper dips in reflection, as the laser spends more time on resonance.

Figures \ref{figure5}(b) and \ref{figure5}(c) shows the intensity of the reflected light from our cavity (after coupling, but before fully optimizing into the fundamental mode) when scanning a laser quickly over three free spectral ranges. The higher order modes are visible with decreasing amplitude and equally spaced in frequency. Having previously calibrated the conversion between the piezo scan amplitude and the change in frequency for our laser, we measure the frequency spacing between the different modes to be 222 MHz, which is the same as the expected value we calculate from Eq. (\ref{eq:nu_mnp}) for our cavity. The transverse modes were imaged by a camera in transmission and show the expected Hermite-Gaussian intensity patterns. If the setup were perfectly cylindrically symmetric, the modes with the same value of $m+n$ would be exactly degenerate in frequency, and the intensity pattern would be the superposition of all degenerate modes. In Fig. \ref{figure5}(a), the ob
served mode for $m+n = 2$ corresponds to a superposition of the three modes shown in Fig. \ref{figure1}(b). The images of other higher-order modes show less perfect superposition due to the combined effect of imperfections in the cavity symmetry, which lift the degeneracy of modes with the same value of $m+n$,\cite{fabre1986} and of a nonsymmetric coupling of the incident beam, which favors some modes.

Figure \ref{figure5} also shows that the signal in reflection when scanning the laser frequency is not the expected quasi-static response with approximately Lorentzian dips at resonance, as sketched in Fig. \ref{figure1}(c). If one zooms on resonance, one observes an asymmetric, oscillating response [Figs.~\ref{figure5}(c) and \ref{figure5}(d)]. Closer inspection shows a chirped oscillation with a decaying envelope. In Appendix \ref{sec: response to frequency sweep}, we investigate this oscillating response in more detail and show how it provides information on the laser scan rate and cavity finesse.

\subsection{Pound-Drever-Hall locking of the laser} \label{sec: PDH locking}

\begin{figure}
    \centering
    \includegraphics{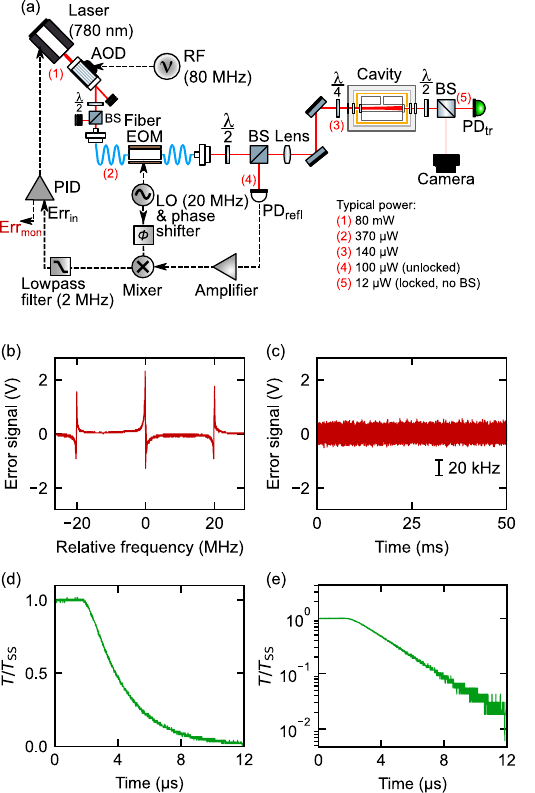}
    \caption{(a) Optical setup for PDH locking of the laser to the cavity. We use a Toptica DL-pro laser that outputs 80 mW and send approximately 350 µW to a fiber EOM (EOSPACE) modulated at 20 MHz by an arbitrary function generator (Tektronix AFG 3022C). The reflection signal from the cavity is sent to a photodiode (ThorLabs PDA10A-EC), and the photodiode signal is amplified (Mini-Circuits ZFL-500+) before being mixed (Mini-Circuits ZAD-1H+) with a phase-shifted version of the modulation signal and sent to the PID controller (Vescent D2-125). The error signal is monitored by an output from the PID and is shown in (b) and (c). For (d) and (e), we use the AOM (AA Opto Electronics at 80 MHz) to abruptly cut the light in the cavity and measure the decay of the transmission signal; $T_{\rm SS}$ is the intensity transmission coefficient in the steady state. Fitting to an exponential decay gives $\tau = 2.6$ $\mu$s.}
    \label{figure6}
\end{figure}

We now turn to locking the laser to our ultrastable cavity using the Pound-Drever-Hall (PDH) technique. In short, the PDH technique generates an error signal with a large capture range. This is achieved by using phase modulation at a frequency $f_{\rm mod}$ to create sidebands on the laser light; when the carrier is not exactly resonant with the cavity, it is partially reflected and beats with the reflected sidebands. The PDH error signal is derived from the amplitude of this beat note, and the capture range is given by $\pm f_{\rm mod}$. For an in-depth explanation of the concepts underlying PDH locking, we refer readers to the excellent introduction in Ref.~\onlinecite{black2000}. In the following, we focus on the practical implementation of the PDH lock. 

Our setup is depicted in Fig. \ref{figure6}(a). As the local oscillator for generating the PDH signal, we use an arbitrary function generator at $f_{\rm mod}=20$ MHz to drive a fiber EOM that modulates the phase of the laser light.\cite{footnote3} The reflection photodiode signal is amplified and mixed with a phase-shifted version of the 20 MHz modulation and then lowpass filtered before being fed into the proportional-integral-derivative (PID) controller. The error signal from the Error Monitor output of our PID controller is shown in Fig. \ref{figure6}(b). We optimize the phase of the local oscillator by first finding the phase that minimizes the error signal peaks and then adding~$90^{\circ}$. We then adjust the DC offset on the PID controller such that the baseline of the error signal is at 0 V. To lock the laser, we slowly decrease the scan amplitude to zero while adjusting the laser frequency to stay centered on the middle error signal peak. We then activate the lock. T
o check whether the laser remains locked on resonance or not, we monitor the signal from the transmission photodiode on the oscilloscope; when locked, the transmission signal should remain near the peak value observed when scanning the laser frequency. 

The PID parameters must then be tuned to optimize the lock. The goal is to maximize the amount of time for which the laser remains locked and to minimize the amplitude of the remaining fluctuations of the error signal. The optimal PID parameters vary for each setup and depend on the parameters of the feedback loop. Manufacturers of PID controllers often have useful information in the provided manuals.\cite{footnote4} Figure \ref{figure6}(c) shows our error signal when the laser is locked, with RMS fluctuations of 0.14 V. Using the slope of the central error signal peak from Fig. \ref{figure6}(b), which is 27 V$/$MHz, we estimate that the RMS frequency fluctuations of our laser are approximately 5 kHz.\cite{footnote5} To further characterize the frequency noise, one can analyze the full noise spectrum of the error signal, as done for instance in Ref.~\onlinecite{deleseleuc2018}.

\subsection{Measuring cavity finesse using the ringdown technique} \label{sec: ringdown finesse measurement}
Once the laser is locked, it is possible to measure accurately the cavity decay time $\tau$ using the ringdown method, which consists in abruptly switching off the incident beam to the cavity and measuring the temporal decay of the transmitted signal. Using an AOM, we shut off the light sent to the cavity in less than 200 ns. Figure \ref{figure6}(d) shows the evolution of the signal in transmission. Fitting the decay by an exponential, we measure a $1/e$ decay constant $\tau = 2.6(1)\,\mu$s, giving a finesse of $2.5(1) \times 10^4$, which is in good agreement with the nominal finesse of the cavity (25,200).

\section{Discussion}
In this article, we have given an introductory overview of both the conceptual background and practical setup of ultrastable cavities for locking lasers. We believe this article will be useful for students and researchers setting up ultrastable cavities for use in atomic and molecular physics experiments. In addition, the material presented here could be adapted for advanced undergraduate instructional laboratory courses to design several projects, such as coupling and mode matching into the cavity, including imaging the higher-order modes (Sec. \ref{sec: mode matching}); locking the laser using the PDH method (Sec. \ref{sec: PDH locking}); measuring the finesse of the cavity using the ringdown technique (Sec. \ref{sec: ringdown finesse measurement}); and investigating the temporal response of the cavity to frequency sweeps in reflection and transmission (Appendix \ref{sec: response to frequency sweep}). Such experiments will give students exposure to many essential concepts 
in modern experimental optics, including Gaussian beam physics, signal modulation, the use of electro-optic and acousto-optic devices, laser locking, and data analysis. We emphasize that for such pedagogical applications, the state-of-the-art equipment described here---such as an ultrastable ULE cavity under vacuum, a fiber EOM, and a high-quality ECDL---is not necessarily required. For instance, it is possible to build a (nonultrastable) cavity of finesse $\simeq 20,000$ using off-the-shelf commercial mirrors and a stainless steel spacer within a budget of a couple thousand dollars,\cite{footnote6} making the experiments described in this article more accessible to students in undergraduate laboratory courses.
 
\begin{acknowledgments}
We thank Renaud Mathevet, Igor Ferrier-Barbut, Gabriel Emperauger, and Bastien Gély for helpful feedback on the manuscript. We also thank Mark Notcutt for permission to use a photo of a Stable Laser Systems cavity. We acknowledge funding from ERC ATARAXIA. 
The authors have no conflicts to disclose.
\end{acknowledgments}

\appendix

\section{Higher-order Gauss-Hermite beams} \label{higher order Gauss-Hermite}
For the sake of completeness, we include here the full expression for the wave amplitudes of higher-order Gauss-Hermite beams in cylindrical coordinates, which is
\begin{align*}
    E_{m,n}(r,z) = & E_0 \frac{w_0}{w(z)} \mathrm{exp} \left( -ikz \right) \times \\
    & H_m \left( \frac{\sqrt{2}x}{w(z)} \right) H_n \left( \frac{\sqrt{2}y}{w(z)} \right) \times \\ & \mathrm{exp} \left[ -r^2 \left( \frac{1}{w^2(z)} + \frac{k}{2R(z)} \right) \right] \times\\
    & \mathrm{exp} \left[ -i(m+n+1) \tan^{-1} \left( \frac{z}{z_R} \right) \right],
\end{align*}
where $w(z) = w_0 \sqrt{1 + \left( {z}/{z_R} \right)^2}$, $k = 2 \pi / \lambda $ is the angular wavenumber, $H_{i}$ are the Hermite polynomials, and the waist is located at $z=0$ \cite{brooker2002}.

\section{Cavity response to a frequency sweep} \label{sec: response to frequency sweep}
\begin{figure}[t]
    \centering
    \includegraphics{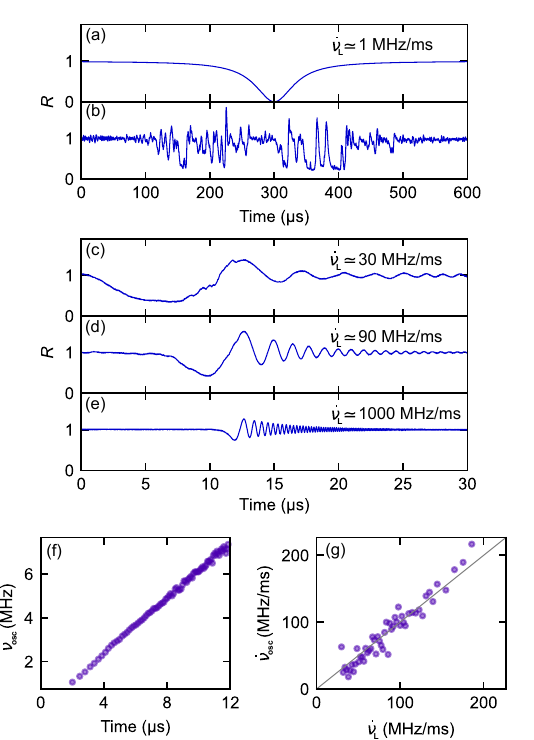}
    \caption{Measured reflection coefficient $R$ for different laser frequency scan rates. (a) Simulation of reflection intensity for $\dot{\nu}_{\rm L} \simeq 1$ MHz/ms. (b) Measured reflection signal for $\dot{\nu}_{\rm L} \simeq 1$ MHz/ms showing the widening of the resonance relative to (a) due to laser jitter. (c, d, e) Measured reflection signals for increasing frequency scan rates, showing that the oscillations increase in frequency with the scan rate. (f) Measured instantaneous oscillation frequency for $\dot{\nu}_{\rm L} \simeq 700$ MHz$/$ms. As shown, the oscillation frequency is a linear chirp whose slope gives the measured value of $\dot{\nu}_{\rm osc}$. (g) Measured values of $\dot{\nu}_{\rm osc}$ for different values of $\dot{\nu}_{\rm L}$, with the gray line representing $\dot{\nu}_{\rm osc}$ = $\dot{\nu}_{\rm L}$.}
    \label{figure7}
\end{figure}

To gain some insight on the origin of the chirped oscillations seen in Fig. \ref{figure5}(d),  we vary the rate $\dot{\nu}_{\rm L}$ at which we scan the laser frequency. Figure \ref{figure7}(a) shows the calculated signal that we would expect at very low scan rates (1 MHz$/$ms) for a laser with zero jitter. Figure \ref{figure7}(b) shows the measured signal in reflection for the same scan rate. Instead of the expected Lorentzian dip, we observe an erratic signal that originates from the laser frequency jumping in and out of resonance with the cavity, as the laser frequency fluctuates erratically on the timescale of tens to hundreds of microseconds during the applied linear scan. 

As we increase the scan rate [Fig.~\ref{figure7}(c) -- \ref{figure7}(e)], we focus on time scales where the laser jitter is less noticeable, and the signal becomes cleaner. As previously noted, the reflection signal shows an oscillatory behavior with a chirp rate that depends on the laser frequency scan rate, $\dot{\nu}_{\rm L}$. To quantify this dependence, we measure the instantaneous frequency $\nu_{\rm osc}$ of the oscillations as a function of time and plot one such measurement (for $\dot{\nu}_{\rm L} \simeq 700$ MHz$/$ms) in Fig. \ref{figure7}(f). From Fig. \ref{figure7}(f), it is clear that the oscillations have a linear chirp, with a slope we denote $\dot{\nu}_{\rm osc}$. Plotting $\dot{\nu}_{\rm osc}$ as a function of $\dot{\nu}_{\rm L}$ [Fig. \ref{figure7}(g)] shows that, to a very good approximation, $\dot{\nu}_{\rm osc} = \dot{\nu}_{\rm L}$: the instantaneous oscillation frequency increases at the same rate as the laser frequency. 

To understand better this behavior, we consider in more detail Fig.~\ref{figure7}(e). At times prior to approximately 11 $\mu$s, the laser is off-resonance, and the intra-cavity field is zero. We thus see the cavity's expected response: all incident light is reflected. Around 12 $\mu$s, the laser becomes resonant with the cavity, and we observe a dip in the reflected signal as we inject light into the cavity. After this initial dip, the incident laser frequency continues to scan away from resonance. We then observe oscillations due to the beating of two fields: the resonant field $E_{\rm cav}$ at frequency $\nu_0$ that has built up within the cavity and is leaking out the first mirror, and the reflected incident laser field $E_{\rm inc}$, whose frequency $\nu_{\rm L}(t)$ is sweeping away from resonance. These two fields beat at their difference frequency: $\nu_{\text{osc}}(t) =  \nu_{\rm L}(t) - \nu_0$, thus explaining the observed relationship $\dot{\nu}_{\rm osc} = \dot{\nu
}_{\rm L}$. We can express the measured intensity in reflection as $I_{\rm ref} \propto |E_{\rm cav} \, e^{-t/2\tau} + E_{\rm inc}|^2$, where $\tau$ is the $1/e$ lifetime of the intracavity intensity. The oscillating term that we observe in the signal arises from the cross-terms and is proportional to $|E_{\rm inc}| \, |E_{\rm cav}| \, e^{-t/2\tau}$. By contrast, in transmission, the measured intensity is given by $I_{\rm tr} \propto |E_{\rm cav} \, e^{-t/2\tau}|^2 = |E_{\rm cav}|^2 \, e^{-t/\tau}$.  Thus, in transmission, we expect a signal that decays as $e^{-t/\tau}$, twice as fast as that in reflection, which decays as $e^{-t/2\tau}$.

\begin{figure}[b]
    \centering
    \includegraphics[width=0.48\textwidth]{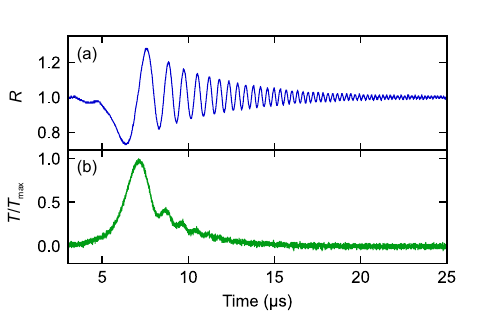}
    \caption{Comparison of signals in reflection (a) and transmission (b). Both signals show chirped oscillations with an overall decaying exponential envelope. Exponential fits of the envelopes give the $1/e$ decay time of the reflection signal as $\tau_{\text{ref}} = 4$ $\mu$s and of the transmission signal as $\tau_{\text{tr}} = 2$ $\mu$s. The signal-to-noise ratio is significantly better in the reflection signal due to heterodyning. Here, we scan the laser frequency at a rate of approximately 300 MHz$/$ms.}
    \label{figure8}
\end{figure}

To confirm this interpretation, we compare in Fig.~\ref{figure8} the reflected and transmitted signals for a scan rate of 300 MHz$/$ms. In transmission [Fig. \ref{figure8}(b)], the signal starts at zero when the laser is off-resonance and then peaks on resonance, as expected. It then decays as the intra-cavity field leaks out. During the decay, the signal exhibits similar oscillations as in reflection, albeit with smaller amplitude. Direct interpretation in terms of beating between different fields is less simple to understand than in reflection, but fitting the decay confirms that it is indeed exponential with time constants in reflection ($\tau_{\rm ref}$) and transmission ($\tau_{\rm tr}$) related by $\tau_{\rm ref} \simeq 2 \tau_{\rm tr}$. For the data in Fig. \ref{figure8}, we find that $\tau_{\rm ref}=4$ $\mu$s and $\tau_{\rm tr}=2$ $\mu$s. The time constant $\tau_{\rm tr}$ is close to the expected value of 2.67 $\mu$s for the cavity decay time $\tau$ calculated from th
e cavity's nominal finesse of 25,200 [see Eq. (\ref{eq:finesse_in_terms_of_tau})]. Measuring the decay constant of the overall envelope in reflection or transmission thus gives a simple way to estimate the finesse of the cavity, though with less accuracy and precision than the (more involved) ringdown technique that we describe in Sec. \ref{sec: ringdown finesse measurement}.

Figure~\ref{figure8} calls for two other observations. First, the reflection signal has a significantly better signal-to-noise ratio than the transmission signal; this is due to heterodyning. The small intracavity field leaking out the first mirror is heterodyned with the relatively large reflected field, which has a much greater amplitude, leading to a better signal quality than the transmitted signal. Additionally, in both reflection and transmission, the change in instantaneous frequency of the oscillations is found to be equal to the laser frequency scan rate. For more quantitative models, see Refs.~\onlinecite{li1991,an1995,poirson1997,lawrence1999}.


\begin{thebibliography}{99}
\bibitem{footnote1}
We neglect variations in the radii of curvature of the mirror because they have a negligible effect relative to $\delta n$ and $\delta L$, as can be calculated from Eq. (\ref{eq:nu_mnp}). For example, the frequency change induced by a 1 $\mu$m change in the length of a 10 cm cavity is on the order of several GHz, whereas a 1 $\mu$m change in the radii of curvature of both mirrors changes the resonance frequency only by several kHz.
\bibitem{footnote3}
A fiber EOM is used in laser-locking setups, as its high bandwidth allows for the creation of sidebands whose frequency $f$ can be tuned all over the free spectral range of the cavity. This makes it possible, when locking such a sideband (say the lower one) on a (fixed) resonance $\nu_{\rm cav}$ of the cavity, to have the laser locked at 
$\nu_{\rm cav}+f$, which can be adjusted at will (e.g. for spectroscopy) by changing the microwave frequency $f$. If frequency tuning of the laser is not needed, one can directly modulate the laser diode current to create the sidebands needed for PDH locking.
\bibitem{footnote5}
Note that this is an in-loop measurement, which only provides a lower bound on the frequency noise. For a more accurate characterization of the noise, an independent ultrastable laser would be needed.
\end{thebibliography}
\end{document}